\documentclass[]{spie}  

\usepackage{amsmath,amsfonts,amssymb}
\usepackage{graphicx}
\usepackage[colorlinks=true, allcolors=blue]{hyperref}

\title{Hybrid Blockchain-Enabled Secure Microservices Fabric for Decentralized Multi-Domain Avionics Systems}

\author[a]{Ronghua Xu}
\author[a]{Yu Chen*}
\author[b]{Erik Blasch}
\author[b]{Alexander Aved}
\author[c]{Genshe Chen}
\author[c]{Dan Shen}
\affil[a]{Binghamton University, SUNY, Binghamton, NY, USA}
\affil[b]{U.S. Air Force Research Laboratory, Rome, NY, USA}
\affil[c]{Intelligent Fusion Tech, Inc, Germantown, MD, USA}

\authorinfo{Further author information: (Send correspondence to Yu Chen)
\\Yu Chen: E-mail: ychen@binghamton.edu
}

\begin{document} 
\maketitle

\begin{abstract}
Advancement in artificial intelligence (AI) and machine learning (ML), dynamic data driven application systems (DDDAS), and hierarchical cloud-fog-edge computing paradigm provide opportunities for enhancing multi-domain systems performance. As one example that represents multi-domain scenario, a ``fly-by-feel'' system utilizes DDDAS framework to support autonomous operations and improve maneuverability, safety and fuel efficiency. The DDDAS ``fly-by-feel'' avionics system can enhance multi-domain coordination to support domain specific operations. However, conventional enabling technologies rely on a centralized manner for data aggregation, sharing and security policy enforcement, and it incurs critical issues related to bottleneck of performance, data provenance and consistency. Inspired by the containerized microservices and blockchain technology, this paper introduces BLEM, a hybrid BLockchain-Enabled secure Microservices fabric to support decentralized, secure and efficient data fusion and multi-domain operations for avionics systems. Leveraging the fine-granularity and loose-coupling features of the microservices architecture, multidomain operations and security functionalities are decoupled into multiple containerized microservices. A hybrid blockchain fabric based on two-level committee consensus protocols is proposed to enable decentralized security architecture and support immutability, auditability and traceability for data provenience in existing multi-domain avionics system. Our evaluation results show the feasibility of the proposed BLEM mechanism to support decentralized security service and guarantee immutability, auditability and traceability for data provenience across domain boundaries.
\end{abstract}

\keywords{Blockchain, Microservices, Dynamic Data Driven Applications Systems (DDDAS), Multidomain Data Analytics, Fly-by-Feel Avionics}

\section{Introduction}
\label{sec:intro}  
As a recent trend, data science has become essential in engineering, business, and medical applications thanks to the advancements in artificial intelligence (AI), machine learning (ML), as well as information fusion technologies \cite{blasch2013revisiting}. Developments in information fusion have moved from surveillance applications based on video and text analytics \cite{hammoud2014automatic, nikouei2019decentralized} towards that of the Internet of things (IoT) scenarios \cite{nikouei2018real, blasch2017panel}, multi-domain applications \cite{rogers2016quest}, and battle management \cite{blasch2016agile}. As an example of multi-domain applications, avionics systems follow principles of layered sensing \cite{mendoza2009video,yang2009performance}, where each layer represents data and information from different domains including space, air, ground, and sea. With the plethora of information available in multi-domain avionics systems, the big data needs to be considered in the 5-V dimensions: volume, velocity, variety, veracity, and value \cite{blasch2013dynamic}. 

As a conceptual framework that synergistically combines models and data in order to facilitate the analysis and prediction of physical phenomena \cite{blasch2018handbook, blasch2018dddas}, DDDAS developments in deep manifold learning \cite{shen2018joint}, nonlinear tracking \cite{yang2005pose, yang2014mobile}, and information fusion \cite{blasch2014static, blasch2014quest, snidaro2016context}, showing promise for advanced avionics assessments. The concept of a DDDAS approach to ``fly-by-feel'' avionics systems is proposed for efficient multi-domain coordination through leveraging modeling (data at rest), real-time control (data in motion) and analytics (data in use) \cite{blasch2019dynamic}. The design of a multi-domain fly-by-feel avionics system could coordinate the space \cite{xu2019exploration}, air \cite{blasch2019blockchain}, ground \cite{blasch2010sensor}, subsurface \cite{wang2011submarine} and cyber domains to determine the mission needs for autonomous surveillance of a designated area.

While DDDAS based ``fly-by-feel'' avionics systems can enhance multi-domain coordination to support multi-intelligence information fusion, it also brings new architecture, performance and security concerns. The multi-domain operations require the coordination among different domain platforms with high heterogeneity, dynamics and different non-standard development technologies. It needs a scalable, flexible and efficient system architecture to support fast development and easy deployment among participants. In addition, to make appropriate, timely decisions in the multi-domain operations, the Android Team Awareness Kit (ATAK) \cite{usbeck2015improving} conveyed Situational Awareness (SA) in a decentralized manner to the users at the edge of the network as well as at operations centers. However, a conventional security and management framework relies on a centralized third-party authority, which can be a performance bottleneck and is susceptible to a single point of failure in distributed SA scenarios, where real-time SA information is shared among geographically scattered command centers and operational troops. Furthermore, DDDAS combines structural health data from the on-board sensors with data from off-line sources for feedback control, Therefore, the data in use should be consistent, unaltered and auditable through the entire lifetime, which means that data quality should be ensured in terms of integrity, traceability and auditability.

In this paper, a hybrid BLockchain-Enabled secure Microservices fabric (BLEM) is proposed to support decentralized, secure and efficient data fusion and multi-domain operations for avionics systems. Leveraging the fine-granularity and loose-coupling features of the microservices architecture \cite{butzin2016microservices,datta2018next}, multi-domain operations and security functionalities are decoupled into multiple containerized microservices. Thus, challenges resulted from the heterogeneity are addressed by allowing development and deployment by participants from different domains, and those lightweight microservices are computationally affordable on resource-constrained IoT devices used in SA scenarios. To enable a decentralized security architecture and support immutability, auditability and traceability for data provenience, a hybrid blockchain fabric is integrated into existing multi-domain avionics concept by using two-level committee consensus protocols. Experimental results demonstrate the feasibility and effectiveness of the proposed BLEM scheme. 

The major contributions of this work are as follows:
\begin{enumerate}
\item A complete architecture of hybrid blockchain-enabled secure microservices fabric for decentralized multi-domain avionics system is proposed, which includes multi-domain fly-by-feel system, secure microservices layer, and a hybrid blockchain network;

\item Security policies, like authentication and access control, are implemented as separate containerized microservices, which utilize a smart contract to act as decentralized application (DApp);

\item A hybrid blockchain fabric, which consists of a two-level consensus protocol, intra-domain consensus and inter-domain consensus, is proposed to improve the scalability and efficiency of consensus in the hierarchical multi-domain network; and

\item A proof-of-concept prototype is implemented and tested on the Ethereum and Tendermint blockchain network, and the evaluation results show that the proposed BLEM scheme provides a decentralized security service and guarantees immutability, auditability and traceability for data provenience in multi-domain scenarios.
\end{enumerate}

The remainder of this paper is organized as follows: Section \ref{sec:relatedwork} reviews background knowledge of DDDAS based multi-domain avionics systems, and the state of the art in blockchain-based decentralized solutions. Section \ref{sec:architecture} illustrates the details of the proposed hybrid blockchain fabric for multi-domain avionics systems. The experimental results and evaluation are discussed in Section \ref{sec:experiment}. Finally, the summary, current limitations and future works are discussed in Section \ref{sec:conclusions}.


\section{State of art and Related Work}
\label{sec:relatedwork}  

\subsection{Dynamic Data Driven Applications Systems (DDDAS)}
\begin{figure} [ht]
\begin{center}
\begin{tabular}{c}
\includegraphics[height=5cm]{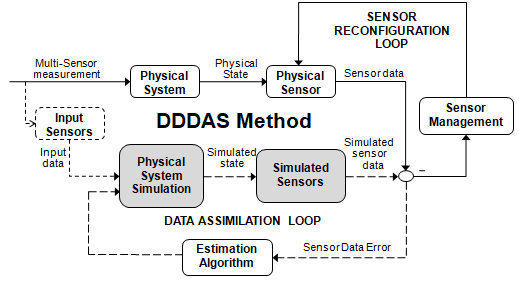}
\end{tabular}
\end{center}
\caption[example] {Dynamic data-driven application systems (DDDAS) concept.\cite{blasch2019study}}
\label{fig:1-DDDAS}
\end{figure}

Dynamic Data Driven Applications Systems (DDDAS) is a conceptual framework that synergistically combines models and data in order to facilitate the analysis and prediction of physical phenomena. In a broader context, DDDAS is a variation of adaptive state estimation that uses a sensor reconfiguration loop as shown in Fig. \ref{fig:1-DDDAS} \cite{blasch2019study}. This feedback loop seeks to reconfigure the sensors in order to enhance the information content of the measurements. The sensor reconfiguration is guided by the simulation of the physical process. Consequently, the sensor reconfiguration is \emph{dynamic}, and the overall process is \emph{data driven}.

The core of the DDDAS is the data assimilation loop, which uses sensor data error to drive the physical system simulation so that the trajectory of the simulation more closely follows the trajectory of the physical system. The \emph{data assimilation loop} uses input data if input sensors are available. The innovative feature of DDDAS paradigm is the additional \emph{sensor reconfiguration loop}, which guides the physical sensors in order to enhance the information content of the collected data. The data assimilation and sensor reconfiguration feedback loops are \emph{computational} rather than physical feedback loops. The simulation guides the sensor reconfiguration and the collected data, and in turn, improves the accuracy of the physical system simulation. The ``model-based simulated data'' positive feedback loop is the essence of DDDAS. Key aspects of DDDAS include the algorithmic and statistical methods that incorporate the measurement data with that of the high-dimensional modeling and simulation. The power of DDDAS is to use simulated data from a high-dimensional model to augment measurement systems for systems design to leverage statistical methods, simulation, and computation architectures \cite{blasch2019dynamic}.

The DDDAS concepts developed over two decades with the simulation methods includes scientific theory, domain methods and architecture design. Scientific theory utilizes modeling and analysis for enhancing the phenomenology of science models by using measurement information and adaptive sampling incorporated into multiphysics, for example avionics\cite{imai2017airplane} and smart cities \cite{fujimoto2016dynamic}. Domain methods utilize data assimilation and multimodal analysis to that of control and filtering for methods of tracking \cite{dunik2015random,jia2016cooperative}, situation awareness \cite{blasch2012wide}, and context-enhanced information fusion \cite{snidaro2016context}. Architecture design is mainly for designing scalable systems architectures and cyber network analysis, with recent efforts in cloud computing based information fusion \cite{liu2014information,wu2017container}.

\subsection{Multi-domain Fly-by-Feel Avionics}
\begin{figure}[ht]
\begin{center}
\begin{tabular}{c}
\includegraphics[height=6.5cm]{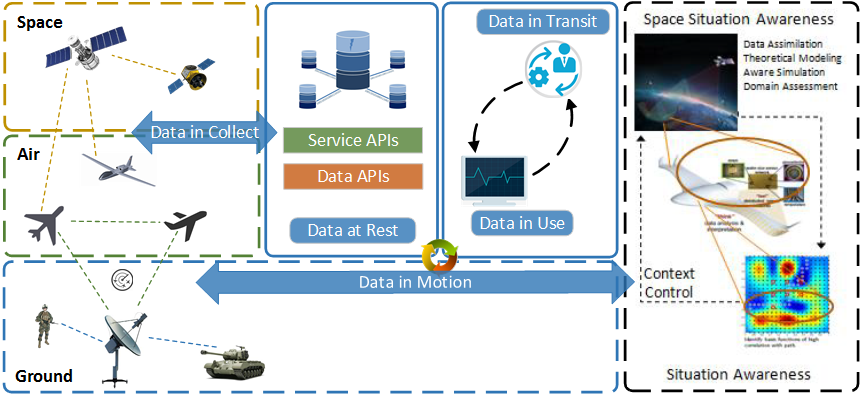}
\end{tabular}
\end{center}
\caption[example] {Multi-domain coordination for fly-by-feel avionics system. }
\label{fig:2-multidomain_system}
\end{figure}

In the fly-by-feel DDDAS approach \cite{kopsaftopoulos2019data}, the structures of the aircraft can provide real-time measurements to adjust the flight control. The integration of on-line data with the off-line model creates a positive feedback loop, where the model judiciously guides the sensor selection, sensor data collection, from which the sensor data improves the accuracy of the flight control model. From the recent \emph{Handbook on Dynamic Data Driven Applications Systems} \cite{blasch2018handbook}, multi-domain scenarios demonstrate techniques to incorporate physics models in support of domain specific operations. Figure \ref{fig:2-multidomain_system} illustrates a multi-domain fly-by-feel concept for future UAVs (or a swarm of UAVs), which leverages DDDAS developments for multi-domain coordination among different platforms in space, air, and ground domains.

\begin{enumerate}
\item \emph{Space Domain}: provides valuable functions for navigation, communication, data routing and services for data in motion. In space situation awareness, space weather detection is important for the continuous satellite operations \cite{shu2016mitigation}, and it can help mitigate the effects of threats to satellites supporting tracking, communication, navigation, and remote sensing \cite{jia2014cooperative,blasch2012orbital}. Current DDDAS developments in situation awareness focus on the results of weather effecting reliable communications \cite{shen2012models,tian2012jamming,wang2016performance}. Satellite health monitoring (SHM) includes the power and electronics to control the satellite \cite{yu2013effectiveness,tian2016joint}. Secure uplink and downlink services can provide data in collect \cite{liu2014adaptive,liu2015real}. The space domain is critical for multi-domain services such as the control and positing of a UAV that provides situation awareness. 

\item \emph{Air Domain}: provides the coordinated autonomous actions on information fusion and control diffusion for data in collect and work as a network of swarm UAVs \cite{chen2007novel}. A recent example of multidomain concept is fly-by-feel that incorporates active sensing for flying \cite{mitchell1995feel}. To enable fly-by-feel concept, various sensors need to be designed \cite{chaney2012pliable} to leverage the other domains such as that of biological systems \cite{salowitz2012bio}. Aeroelastic sensing \cite{mangalam2014fly,suryakumar2015control}, is evident as a DDDAS method to enhance real time management and control in fly-by-feel system. The fly-by-feel techniques incorporate stochastic sensing and filtering as part of the on-line structural health of the aircraft that is incorporated with the measurements of position and air fluid flow \cite{kopsaftopoulos2016stochastic,armanious2017fly}.

\item \emph{Ground Domain}: The Android Team Awareness Kit (ATAK) \cite{usbeck2015improving} is a situation awareness tool that includes many feature displays for a portable device that supports multi-domain operations. ATAK focuses on improving the real-time SA of small units at the tactical edge. which means knowing where you are, where the rest of your team is, and having a variety of ways to communicate with your team (and, if feasible with reach-back, to operation centers) \cite{usbeck2015improving}. While ATAK features the display of various data sources, for multidomain operations; it could provide additional information to the user towards the health of the systems for command and control \cite{blasch2014decisions}. The DDDAS rendering options support the design of a User Defined Operating Picture (UDOP) \cite{blasch2013enhanced} that can be displayed on the ATAK system. The ability to plot tracks, discussions, and labels of objects \cite{connare2001group,blasch2004ten} enhances the situation understanding \cite{blasch2006issues,blasch2012high}.
\end{enumerate}

As Fig. \ref{fig:2-multidomain_system} shows, multi-domain operations require cross-domain data sharing techniques include: data in collect, data at rest, data in use, data in transit and data in motion. 
\emph{Data at Rest} acts as long-term storage service which provides structure (i.e., translations) between data for integration, analysis, and storage.
\emph{Data in Collect} leverage the power of modeling from which data is analyzed for information, delivered as knowledge, and supports prediction of data needs.
\emph{Data in Transit} works as a Data as a Service (DaaS) architecture that incorporates contextual information, metadata, and information registration to support the systems-of-systems design.
\emph{Data in Motion} utilizes feedback control loops to dynamically adapt to changing priorities, timescales, and mission scenarios.
The intersection of the information is \emph{Data in Use}, which provide context-based human-machine interactions based on dynamic mission priorities, information needs, and resource availability. 

\subsection{Microservices in IoT}
The traditional service-oriented architecture (SOA) utilizes a monolithic architecture that constitutes different software features in a single interconnected and interdependent application and database. Owing to the tightly coupled dependence among functions and components, such a monolithic framework is difficult to adapt to new requirements in an IoT-enabled system, such as scalability, service extensibility, data privacy, and cross-platform interoperability \cite{datta2018next}. Though encapsulating a minimal functional software module as a fine-grained and independently executable unit, the \textit{microservices architecture} allows for fast development and easy deployment in multi-domain scenarios. The individual microservices communicate with each other through a lightweight and asynchronous manner, such as HTTP RESTful API. Finally, multiple decentralized individual microservices cooperate with each other to perform the functions of complex systems. The flexibility of microservices enables continuous, efficient, and independent deployment of application function units. As two most significant features of the microservices architecture, \textit{fine granularity} means each of the microservices can be developed in different frameworks and with minimal development resources, while \textit{loose coupling} implies that functions of microservices and its components are independent of each other's deployment and development \cite{yu2018survey}.

Thanks to the fine-granularity and loose-coupling properties, the microservices architecture has been investigated in many smart developments to improve the scalability and security of IoT-based applications. The IoT systems are advancing from ``things''-oriented ecosystem to a widely and finely distributed microservices-oriented-ecosystem \cite{datta2018next}. To enable a more scalable and decentralized solution for advanced video stream analysis for large volumes of distributed edge devices, a system design of a robust smart surveillance systems was proposed based on microservices architecture and blockchain technology \cite{nagothu2018microservice, nikouei2019decentralized, xu2019blendmas}. It aims at offering a scalable, decentralized and fine-grained access control solution for smart public safety. A BlendSM-DDM \cite{xu2019blendsm} is proposed by decoupleing business logic functions and security services into multiple containerized microservices rather than using a monolithic service architecture, and it supports loose-coupling, fine-granularity and easy-maintenance for decentralized data marketing applications.

\subsection{Blockchain and Smart Contract}
As a fundamental technology of Bitcoin \cite{nakamoto2019bitcoin}, \textit{blockchain} initially was used to promote a new cryptocurrency that performs commercial transactions among independent entities without relying on a centralized authority, like banks or government agencies. Essentially, the blockchain is a public ledger based on consensus rules to provide a verifiable, append-only chained data structure of transactions. Blockchain relies on a decentralized architecture which data is verified, stored and updated distributively. In a blockchain network, a \textit{consensus mechanism} is enforced on a large amount of distributed nodes called miners to maintain the sanctity of the data recorded on the blocks. The transactions are validated by miners and recorded in the time-stamped blocks, and each block is identified by a cryptographic hash and chained to preceding blocks in a chronological order.  Thanks to the “trustless” consensus protocol running on miners across the network, participants can trust the system of the public ledger stored worldwide on many different decentralized nodes maintained by ''miner-accountants'', as opposed to having to establish and maintain trust with a transaction counter-party or a third-party intermediary \cite{swan2015blockchain}. Thus, blockchain offers a prospective decentralized architecture to support secure distributed transactions among all participants in a trustless multidomain environment,

Emerging from the intelligent property, a \textit{smart contract} allows users to achieve agreements among parties through a blockchain network. By using cryptographic and security mechanisms, a smart contract combines protocols with user interfaces to formalize and secure relationships over computer networks \cite{szabo1997formalizing}. A smart contract includes a collection of pre-defined instructions and data that have been saved at a specific address of blockchain as a Merkle hash tree, which is a constructed bottom-to-up binary tree data structure. Through exposing public functions or application binary interfaces (ABIs), a smart contract interacts with users to offer the predefined business logic or contract agreement.

The blockchain and smart contract enabled security mechanism for applications has been a hot topic and some efforts have been reported recently, for example, smart surveillance system \cite{nagothu2018microservice, nikouei2018real, xu2019blendmas}, social credit system \cite{xu2018constructing}, decentralized data marketing \cite{xu2019blendsm,ramachandran2019trinity}, space situation awareness \cite{xu2019exploration}, biomedical imaging data processing \cite{xu2019decentralized}, and access control strategy \cite{xu2018blendcac, xu2018smartcac}. Blockchain and smart contract together are promising to provide a decentralized solution to support secured data sharing and accessing in multi-domain avionics systems.

\section{BLEM System Architecture}
\label{sec:architecture}
The design of a multi-domain fly-by-feel avionics system requires operation coordination and data exchange across boundaries of space, air, ground and the cyber domain. Such a multi-domain system is deployed in a heterogeneous network environment that with high dynamics and different technologies. In addition, advancement in edge computing based SA, like ATAK, also requires a lightweight and scalable architecture to enable services on a large volume of resource constrained IoT devices. The virtualization technology, like virtual machines (VMs) or containers, is platform independent and could provide resource abstraction and isolation features, they are ideal for system architecture design to address the heterogeneity challenge in multi-domain scenarios. Compared to VMs, containers are more lightweight and flexible with operating system (OS)-level isolation, so that is an ideal selection for service deployment on edge computing platforms. 

Widely used ATAK technology can improve accuracy and real-time decision for multi-domain task through a decentralized SA manner. However, existing security and management frameworks normally rely on a centralized authority, which can be a performance bottleneck or susceptible to a single point of failure. Furthermore, cross-domain data sharing technologies is essential for DDDAS operations like feedback control, so that the data should be consistent, unaltered and auditable through the entire lifetime. To address above issues, blockchain and smart contract offer a promising solution to enable a decentralized trust network and secure data sharing service, where data and its history are reliable, immutable and auditable.

\begin{figure}[ht]
\begin{center}
\begin{tabular}{c}
\includegraphics[height=10.5cm]{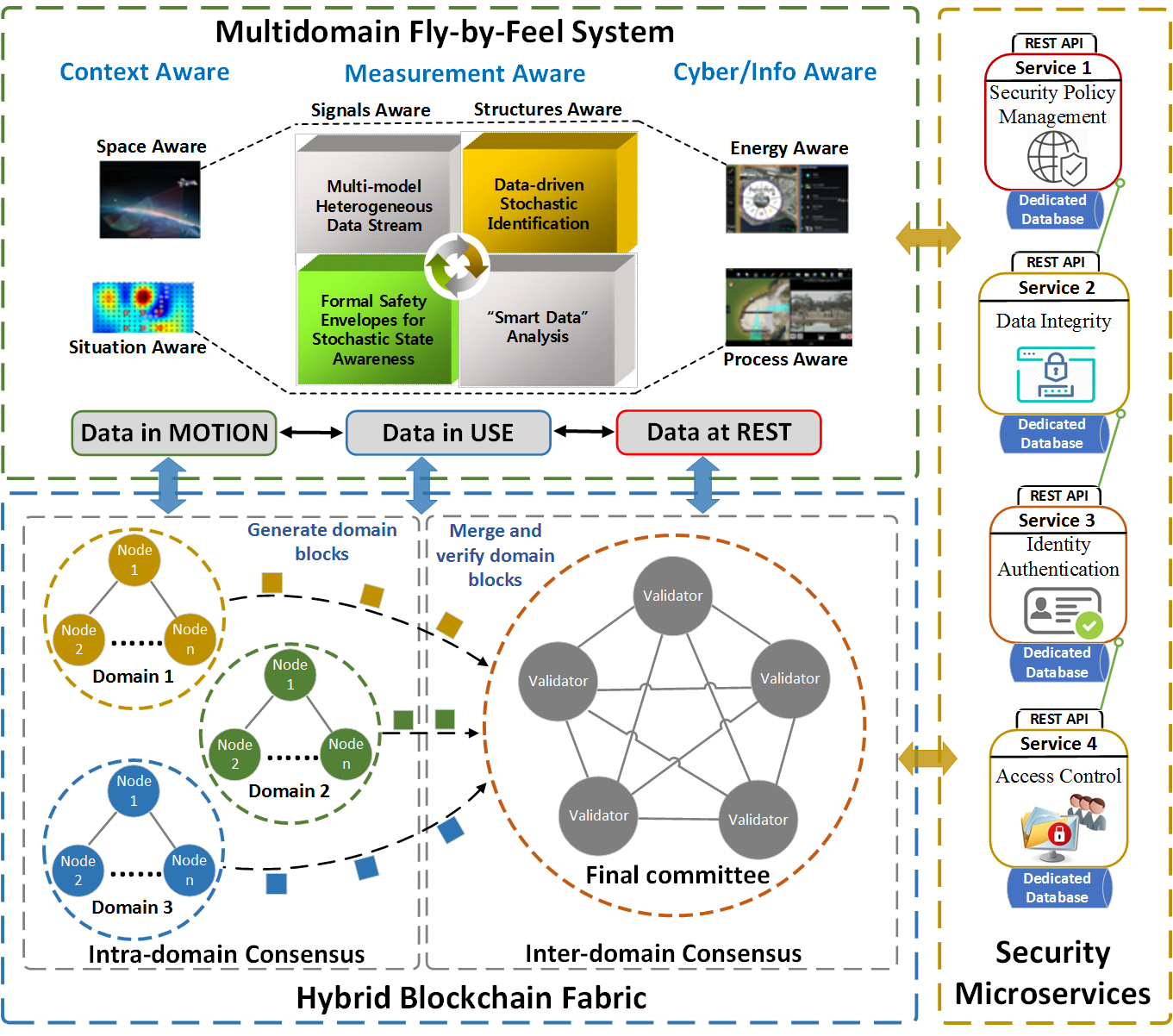}
\end{tabular}
\end{center}
\caption[example] {Architecture of BLEM: a Hybrid Blockchain Fabric for Multi-domain Fly-by-Feel Avionics.}
\label{fig:3-system_architecture}
\end{figure}

Figure \ref{fig:3-system_architecture} illustrates the system architecture of the proposed BLEM scheme, a hybrid blockchain-enabled fabric for multi-domain fly-by-feel avionics system. The whole system consists of (i) a multi-domain fly-by-feel system that relies on DDDAS method to increase maneuverability, safety and fuel efficiency in avionics scenario, (ii) a blockchain-enabled security services layer that leverages microservices and smart contract to support flexible, efficient and secure multidomain operations, and (iii) a hybrid blockchain fabric as the fundamental network infrastructure that utilizes lightweight consensus protocols and distributed ledger to enable decentralized security mechanism.

\subsection{Multi-Domain Fly-by-Feel System}
The multi-domain fly-by-feel avionics system measures the aerodynamic forces (wind, pressure, temperature) for physics-based adaptive flight control to increase maneuverability, safety and fuel efficiency. The upper left of Fig. \ref{fig:3-system_architecture} presents a DDDAS method that identifies safe flight operation platform position needs from which models, data, and information are invoked for effective flight control. Context, measurement and cyber/info awareness are three methods to support a combined systems awareness analysis. 

\begin{enumerate}
\item Measurement awareness includes signal and structure awareness based on air, fluid, and structural analysis. For structure aware, structures of the aircraft can provide real-time measurements, such as stain and temperature,  to adjust the flight control. Given the data collected by the sensors, signal aware can provide estimates of initial conditions, boundary conditions, inputs, parameters, and states to enhance the accuracy of the model. 

\item Context awareness methods includes space and situation awareness. The space awareness generally consists of two major areas: satellite operations and space weather. The satellite operations are focused on the local perspective to enable continuous operations by understanding the space environment and build models to support satellite health monitoring (SHM) \cite{xu2019exploration}. For context situation awareness, target tracking, pattern classification, and coordinated control are components of information fusion which can applied to video tracking and wide area motion imagery.

\item Cyber/info awareness uses security, power, and scene (data) modeling of the system to enable energy and process awareness. These functions operate over the layered domain operations as DDDAS-based resilient cyber battle management services.
\end{enumerate}

The above fly-by-feel air platform concept leverages modeling (data at rest), real-time control (data in motion) and analytics (data in use) for multi-domain coordination. Given information gathered from space (e.g., GPS), air (e.g., aircraft measurements), and ground Automatic Dependent Surveillance Broadcast (ADS-B), the DDDAS system based on multi-domain coordination can determine the mission needs for autonomous surveillance of a designated area.

\subsection{Security Microservices}
The blockchain-enabled security services layer, as shown in right part of Fig. \ref{fig:3-system_architecture}, acts as a fundamental microservices oriented infrastructure to support decentralized security mechanism. The key elements and operations are described below.

\begin{enumerate}
\item \emph{Service Policy Management}: acts as security service managers who is responsible for entity registration and smart contract authorization. To join the network, a participant uses its blockchain address as request to entity registration process which associates entity’s unique blockchain account address with a Virtual ID (VID) \cite{xu2019exploration}. For smart contract authorization, domain owners or system administrator deploy the smart contracts that encapsulate security function, like data integrity and access control. After the smart contracts have been deployed successfully on the blockchain network, only authorized participants could interact with smart contract through the Remote Procedure Call (RPC) interfaces.

\item \emph{Data Integrity}: to support DDDAS multidomain task, data fusion among online (data in motion) and offline (data at rest) is need and intersection of the information is data in use. Thus, it necessary to ensure data integrity as combining those data in decision-making tasks. Data integrity technologies are mainly to ensure reliable and immutable data access at the same time avoid storing a huge amount of redundant data in the blockchain. The data integrity microservices provides the dynamic data synchronization and efficient verification through a hashed index authentication process by smart contract \cite{nikouei2018real}. The data owners just simply save the hashed index of data to distributed ledger through authorized ABI functions of smart contract. In verification process, data user just fetch a key-value index from distributed ledge and compares it with calculated hash values of the received data.

\item \emph{Identity Authentication}: Since each blockchain account is uniquely indexed by its address that is derived from his/her own public key, the account address is ideal for identity authentication needed by other security microservices, such as data integrity and access control. Once an identity verification service request is acknowledged, the identity authentication decision making process checks the requester identity profile by referring with the RESTful API to other microservices-based service providers for referring identity verification results.

\item \emph{Access Control}: The domain administrator and data owners could transcode access control (AC) models and policies into a smart contract-based access control (AC) microservice \cite{xu2018blendcac, xu2018smartcac, xu2019blendmas}. To successfully access data or execute task in multidomain coordination, an user initially sends an access right request to the AC microservices to get a capability token. Given results from identity verification and access right decision making process, the AC microservice issues the capability token encoding authorized access right and update the token data in the smart contract.

\end{enumerate}

The security microservices allows service providers and data owner to deploy their own security policies as smart contracts instead of relying on a centralized third party authority. It provides a decentralized security mechanism for distributed multi-domain scenarios.

\subsection{Hybrid Blockchain Fabric}

The hybrid blockchain fabric is responsible for consensus protocol and persistent storage, which are enabling technology for decentralized security mechanism. As the core of blockchain, the \emph{consensus protocol} is mainly to maintain data integrity, consistence and order of data in the distributed ledger across the trustless multi-domain network. To improve the scalability and efficiency of executing consensus protocols in a multi-domain network with heterogeneity and dynamics, a two-level consensus protocol is proposed: intra-domain consensus and inter-domain consensus, as shown at the bottom of Fig. \ref{fig:3-system_architecture}. For an individual domain, a classical Byzantine Fault Tolerant (BFT) \cite{lamport2019byzantine} based intra-committee consensus protocol is executed among committee member to validate a disjoint set of transactions within domain. For multi-domain coordination, an inter-domain consensus protocol is responsible to validate those blocks across domain boundary and finalize a global distributed ledger. Key components and workflows are explained as follows:

\begin{enumerate}
\item \emph{Permissioned committee network}: Following the idea of delegation, only a small subset of the nodes in the network are selected as validators who form a committee and perform the consensus protocol. Permissioned networks provide basic security primitives, such as public key infrastructure (PKI), identity authentication and access control, etc. Public key cryptography is used to secure communication and transactions validation, like digital signature, etc. 

\item \emph{Intra-domain consensus}: The BFT replication consensus protocols, like Practical BFT (PBFT) \cite{castro1999practical}, execute the consensus algorithm among a small group of nodes which are authenticated by the network administrator. They are well adopted in the permissioned blockchain network in which the access control strategies for network management are enforced. For each domain, data transactions within domain are broadcasted among validators who record verified transactions in blocks. The consensus
agreement is achieved as those proposed intra-domain blocks are signed by no less than 2/3 of validators in the committee. Owing to the small size of the intra-domain committee, only a limited network delay is introduced for messages propagation, so that it ensures high throughput of transactions in intra-domain scenarios, which require high data transactions rate and fast response to service requests.

\item \emph{Inter-domain consensus}: To jointly address several critical issues such as pseudonymity, scalability and poor synchronization in an open-access inter-domain network environment, the Proof-of-Concept (PoC) consensus mechanism, like PoW, is adopted as the inter-domain consensus protocol. The inter-domain committee is responsible to verify data transactions across inter-domain, and propose new block containing verified transactions, then finalize blocks in a global distributed ledger. The security of the consensus protocol requires that the majority (51\%) of the nodes are honest and they can correctly execute the consensus protocol. The inter-domain consensus is aimed to support the scalability and probabilistic finality in the partial synchronous multi-domain networks environment.

\end{enumerate}

\section{Implementation and Evaluation}
\label{sec:experiment}
To verify the proposed BLEM scheme, a proof-of-concept prototype is implemented in a real physical network environment. The security microservices have been implemented as Docker containers, which are deployed both on the edge (Raspberry Pi) and fog (desktop) units. The web service application development is built on Flask framework \cite{flask} using Python. For the blockchain part, we use Ethereum \cite{ehtereum} for inter-domain operations, while Tendermint \cite{kwon2014tendermint} is used for intra-domain consensus mechanism. The smart contract development use Solidity \cite{solidity}, which is a contract-oriented, high-level language for implementing smart contracts.

\subsection{Experimental Setup}
Table \ref{tab:testbed} shows configurations of nodes used in the experiments. In this prototype, the laptops acts as domain administrators, which takes role of oracle to manage domain network. All desktops work as fog computing nodes, while a Raspberry PI runs as edge computing node. The inter-domain network is built on a Ethereum private network which includes six desktops as miners and two Raspberry PIs as nodes. The security microservices are hosted both on fog and edge computing nodes. All devices use Go-Ethereum \cite{goethereum} as the client application to interact with ethereum network. The intra-domain network is built on a private Tendermint network which uses 16 Raspberry PIs as validators.

\begin{table}[ht]
\caption{Configurations of Experimental Nodes.} 
\label{tab:testbed}
\begin{center}       
\begin{tabular}{|l|p{5.0cm}|p{7.5cm}|} 
\hline
\rule[-1ex]{0pt}{3.5ex} \textbf{Device} & Dell Optiplex 760 & Raspberry Pi 3 Model B+ \\
\hline
\rule[-1ex]{0pt}{3.5ex} \textbf{CPU} & 3 GHz Intel Core TM (2 cores) & Broadcom ARM Cortex A53 (ARMv8), 1.4GHz \\
\hline
\rule[-1ex]{0pt}{3.5ex} \textbf{Memory} & 4GB DDR3 & 1GB SDRAM \\
\hline
\rule[-1ex]{0pt}{3.5ex} \textbf{Storage} & 250G HHD & 32GB (microSD card) \\
\hline
\rule[-1ex]{0pt}{3.5ex} \textbf{OS} & Ubuntu 16.04 & Raspbian GNU/Linux (Jessie) \\
\hline
\end{tabular}
\end{center}
\end{table}
\vspace{-10 pt}

\subsection{Performance Evaluation}
To evaluate the performance of the microservices-based security mechanism, a service access experiment is carried out on a physical network environment by simulating service request and acknowledge. A Raspberry PI works as a client to send service request, while server side is a service provider, who has been both hosted on Raspberry Pi (edge) and Desktop (fog) nodes. For blockchain fabric evaluation, we focus on transaction rate and throughput by calculating transactions committed time on Tendermint network.

\subsubsection{Security Service Overhead}

\begin{figure}[ht]
\begin{center}
\begin{tabular}{c}
\includegraphics[height=6.0cm]{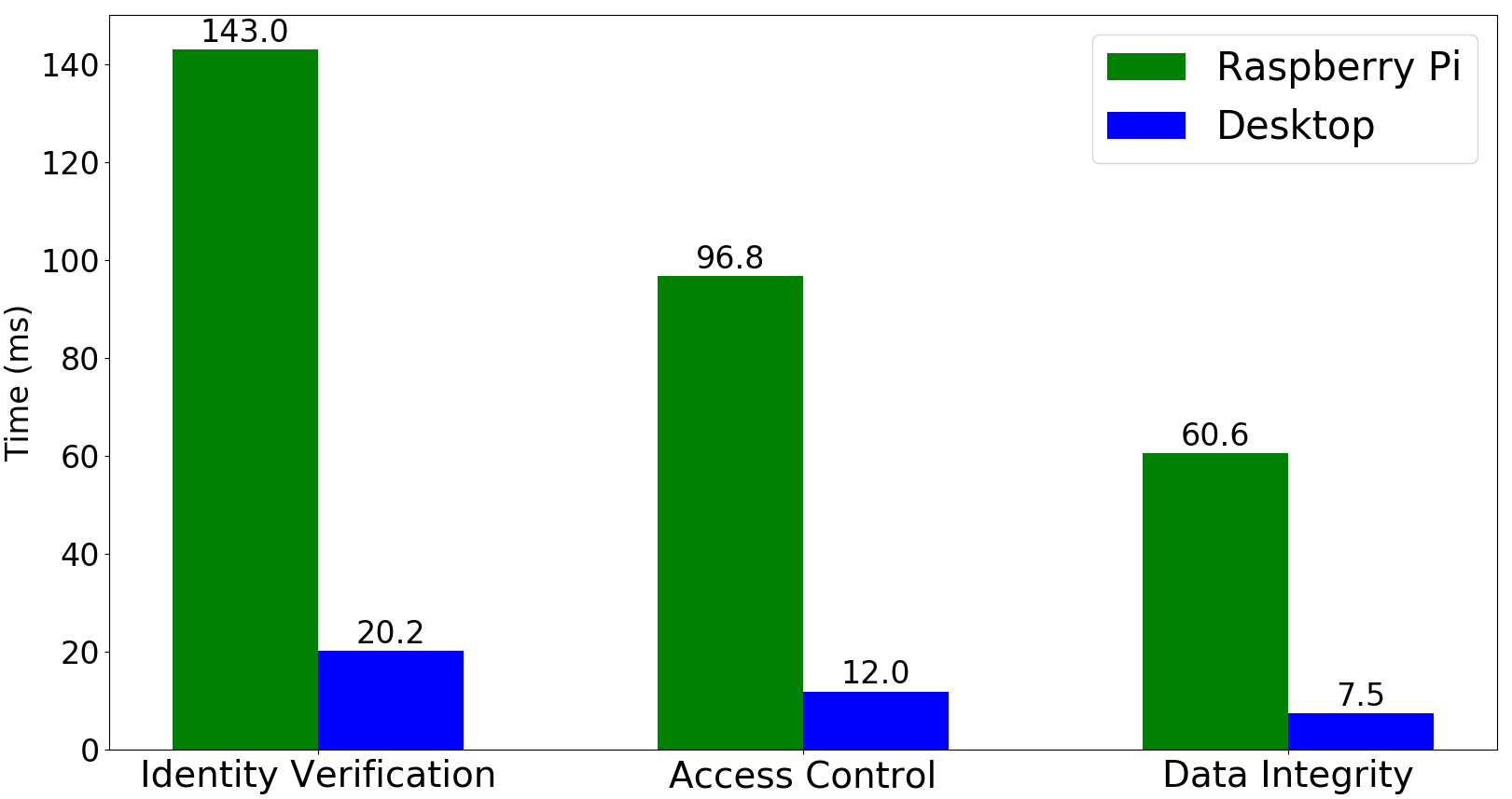}
\end{tabular}
\end{center}
\caption[example] { Performance of running security microservices. }
\label{fig:4-service_performance}
\end{figure}

To evaluate the overhead of running microservices on the host machine, key security microservices including identity verification, access control and data integrity microservices are deployed on three Raspberry Pi and three desktops, separately. 50 test runs have been conducted based on the proposed test scenario, where the client sends a data query request to server side for an access permission. Figure \ref{fig:4-service_performance} demonstrates the computation overhead incurred by running individual microservice on different platform. The results show that computation overhead increase as the task complexity grows. 

Compared with data integrity, access control and identity verification consist of more cryptography and authentication operations. Therefore, they incur higher computation overhead both on the Raspberry Pi and the desktop. Since identity verification microservice involves multiple smart contract interactivities, like registry reference and identity authentication, it takes longer execution time for querying the data in blockchain.

\subsubsection{Network Latency}
\begin{figure}[ht]
\begin{center}
\begin{tabular}{c}
\includegraphics[height=6.4cm]{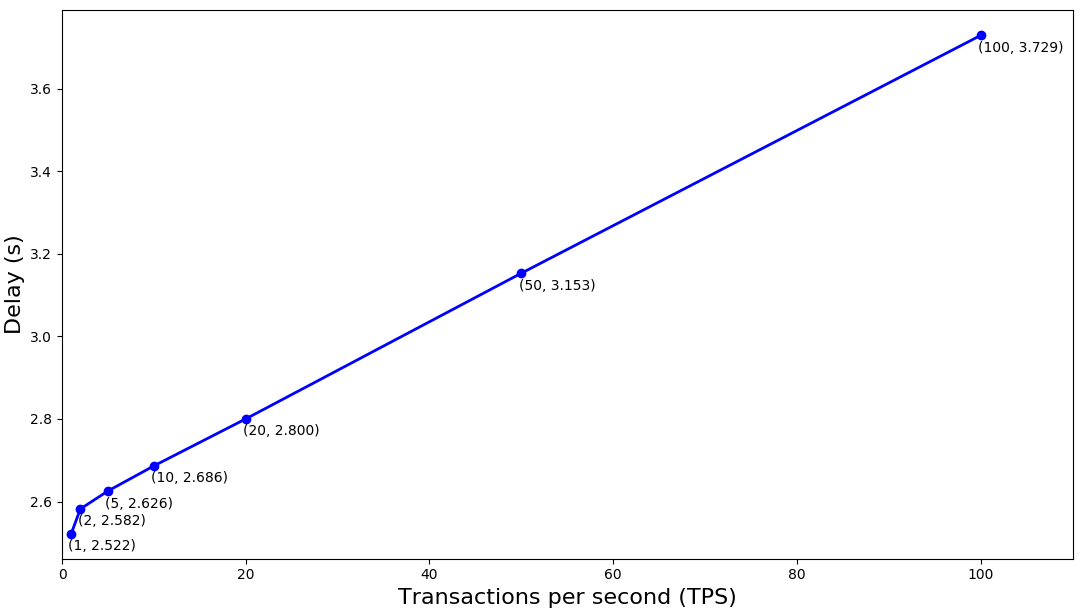}
\end{tabular}
\end{center}
\caption[example] { Delay with different transaction rate. }
\label{fig:5-tx_rate}
\end{figure}

For an intra-domain committee, validators receive and verify transactions, and execute BFT consensus to guarantee security of the distributed ledger. The consensus protocol and ledger storage process inevitably introduce extra delays on normal service requests and operations. Figure \ref{fig:5-tx_rate} shows the network latency when a validator publishes a transaction within the domain and waits until it committed on the ledger. The network latency is measured by committing fixed size transaction data in domain committee given difference transaction rate. The transaction used in the test is 1 KB to reduce the influence of data size on network performance. Given test Tendermint network with 16-validator Raspberry Pi devices, we evaluated the end-to-end delay with a validator sending multiple transactions per second (TPS), which varies from one to 100 TPS. In terms of the communication complexity of broadcasting transactions, the latency of committing transactions is almost linear scale to the transaction rate, and it varies from 2.5 s to 3.7 s. For the inter-domain scenario, sixty blocks were appended to the blockchain and the average block confirmation time was calculated as 7.7 s on our Ethereum private network.

\subsubsection{Throughput Evaluation}
\begin{figure}[ht]
\begin{center}
\begin{tabular}{c}
\includegraphics[height=6.8cm]{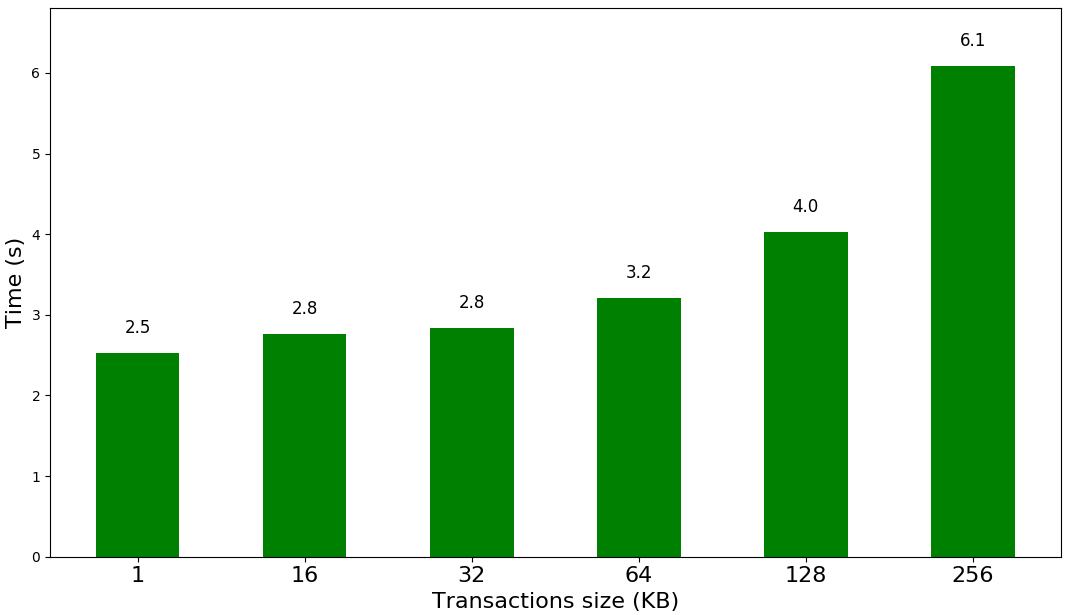}
\end{tabular}
\end{center}
\caption[example] { Throughput evaluation. }
\label{fig:6-tx_size}
\end{figure}

Figure \ref{fig:6-tx_size} shows the time that takes for an intra-domain committee to complete an entire consensus protocol run with variable transaction size between 1K and 256K. The transaction rate in this test is 1 TPS to reduce the influence of data traffic on network performance. The transaction data throughput is specified in M/h, means Mbytes per hour. With variant data sizes, corresponding results are obtained as shown in Table \ref{tab:throughput}. Given a fixed transaction rate of 1 TPS, increasing the transaction size allows committing more data on the distributed ledger, and therefore reach a higher throughput, which maximizes the system capability.

\begin{table}[t]
\caption{Data Throughputs vs. Transaction Data Sizes.} 
\label{tab:throughput}
\begin{center}       
\begin{tabular}{|l|c|c|c|c|c|c|} 
\hline
\rule[-1ex]{0pt}{3.5ex} \textbf{Transaction Size} & 1K & 16K & 32K & 64K & 128K & 256K \\
\hline
\rule[-1ex]{0pt}{3.5ex} \textbf{Throughput (M/h)} & 1.4 & 20.9 & 40.6 & 71.9 & 114.5 & 151.5 \\
\hline
\end{tabular}
\end{center}
\end{table}

\section{Conclusions}
\label{sec:conclusions}
In this paper, BLEM, a hybrid blockchain-enabled secure microservices fabric is proposed to enable decentralized security mechanism and support secure and efficient data fusion and multi-domain operations for multi-domain avionics system. A comprehensive overview of the system architecture is presented, and critical elements are illustrated. A concept-proof prototype has been developed and verified on a physical network environment. The experimental results demonstrate the feasibility of proposed solutions to address performance and security issues in multi-domain avionics systems.

While the reported work has shown great potential, there is still open questions to be addressed before a practical decentralized security solution can be deployed in real-world multi-domain avionics application. Future efforts include further simulation and development towards a prototype for multi-domain avionics scenarios.

\bibliography{report} 
\bibliographystyle{spiebib} 

\end{document}